# Effects of hydrodynamic noise on the diffusion of polymers in dilute solutions

**V. Lisy** [a] [1], **J. Tothova** [b], and **A.V. Zatovsky** [c]

[a] Department of Physics, Technical University of Kosice, Park Komenskeho 2, 042 00 Kosice, Slovakia
[b] Institute of Physics, P. J. Safarik University, Jesenna 5, 041 54 Kosice, Slovakia
[c] Department of Theoretical Physics, I. I. Mechnikov Odessa National University, 2, Dvoryanskaya Str., 65026 Odessa, Ukraine

E-mail: vladimir.lisy@tuke.sk, jana.tothova@upjs.sk

**Abstract.** The Rouse-Zimm equation for the position vectors of beads mapping the polymer is generalized by taking into account the viscous aftereffect and the hydrodynamic noise. For the noise, the random fluctuations of the hydrodynamic tensor of stresses are responsible. The preaveraging of the Oseen tensor for the nonstationary Navier-Stokes equation allowed us to relate the time correlation functions of the Fourier components of the bead position to the correlation functions of the hydrodynamic field created by the noise. The velocity autocorrelation function of the center of inertia of the polymer coil is considered in detail for both the short and long times when it behaves according to the $t^{-3/2}$ law and does not depend on any polymer parameters. The diffusion coefficient of the polymer is close to that from the Zimm theory, with corrections depending on the ratio between the size of the bead and the size of the whole coil.

**Keywords:** Flexible polymer, Rouse-Zimm model, dilute solution, fluctuating hydrodynamics

---

[1] Author to whom any correspondence should be addressed.

________________________________________________________________

**Contents**

________________________________________________________________

**Introduction**

    The most popular theories of polymer dynamics are the bead-spring Rouse [1] and Zimm [2] phenomenological models that are considered universal models for the polymer long-time dynamics [3, 4]. In spite of some fifty years of investigations of these models, there are still problems in their applications in the interpretation of experiments, such as the dynamic light or neutron scattering [5 - 7]. For example, the "universal" plateau in the Zimm plot of the first cumulant of the dynamic structure factor is lower than predicted in the Rouse-Zimm (RZ) theory for flexible polymers. The diffusion coefficients of the polymer coils determined from the dynamic scattering at small wave vector transfers do not correspond to the static scattering data, etc. The origin of these discrepancies remains unclear for decades [8, 9]. The development of the theory of polymer dynamics, even in the simplest case of ideal flexible polymers in dilute solutions, is thus still of interest.

    In our previous papers [10, 11], a generalization of the RZ theory that can be called the hydrodynamic RZ model has been proposed. The main idea of this generalization comes from the theory of the Brownian motion, which lies in the basis of the RZ model. In the usual (Einstein) description [3, 4], the resistance force on the bead during its motion in the solvent is the Stokes force, which is at a given moment of time $t$ determined by the bead velocity at the same time. This approximation is valid only for the steady motion, i.e., at $t \to \infty$. In a more general case, within the Navier-Stokes hydrodynamics, the friction force on the Brownian particle should be the Boussinesq force [12] called the history force since at the time $t$ it depends on the state of the particle motion in all the preceding moments of time (for incompressible fluids $t \gg b/c$ where $c$ is the velocity of sound and $b$ the radius of the spherical particle). This was first noted by Vladimirsky and Terletzky [13] who have built the hydrodynamic theory of the translational Brownian motion. For compressible fluids it was generalized in the work [14], and the hydrodynamic rotational Brownian motion was first considered in Ref. [15]. The hydrodynamic memory (or the so-called viscous aftereffect) is a consequence of fluid inertia. In the Brownian motion it displays in the "long-time tails" in the particle velocity autocorrelation function (VAF) that became famous after their discovery in the computer experiments on simple liquids (for a review see, e.g., [16]). The tails reflect strong correlations with the initial state of the particle and persist for a long time. The time dependence of the mean square displacement (MSD) of the particle changes from the "ballistic" regime at short times to the Einstein diffusion when the MSD is proportional to $t$. The nondiffusive regime with the characteristic time $\tau_b = b^2\rho/\eta$ ($\rho$ is the density of the solvent and $\eta$ its viscosity) was observed in the dynamic light scattering (DLS) experiments, e.g. [17] (for recent experiment, using optical trapping interferometry, see [18, 19]). The sizes of the

studied particles (hundreds of nanometers or even much smaller) were similar to those of long polymer coils in solution. One could thus expect that the hydrodynamic memory affects the dynamics of polymers as well, and that these effects are experimentally detectable.

In the next section we briefly summarize the approach used in our previous papers [10, 11, 20]. Then we derive the generalized RZ equation taking into account the effects of hydrodynamic noise, with random fluctuations of the hydrodynamic stress tensor being responsible for the noise. The spectral properties of the random forces acting on the polymer beads are determined by the hydrodynamic susceptibility of the solvent. Using the preaveraging approximation, we relate the time correlation functions of the Fourier components of the bead radius vector to the correlation functions of the hydrodynamic field created by the noise. The VAF of the center of mass of the coil has been considered in detail. At long times its behavior follows the algebraic $t^{-3/2}$ law and does not depend on the polymer parameters. This particular result exactly corresponds to that known from the theory of Brownian motion for individual rigid particles. It also agrees with the computer simulation study [21] and older theoretical results that take into account the inertial effects in the motion of macromolecules in solution [22, 23]. In the work [22], the macromolecule was regarded as a stiff spherical particle permeable to solvent. Its interaction with the fluid was described by the Brinkman [24] (or Debye-Bueche [25]) equations. In Ref. [23], a similar approach to that used in the present work has been applied, i.e., the random sources have been incorporated in the Navier-Stokes equations to account for the hydrodynamic fluctuations. The authors have obtained equations of motion for the macromolecule in solution and analyzed the influence of nonstationary fluid motion on the dynamics of Zimm polymers through time-dependent correlation functions. However, there are several differences in [23] from our results, which will be discussed below.

**1. Hydrodynamic Rouse-Zimm model**

In the traditional RZ model [3], the forces acting on the $n$th bead are $\vec{f}_n^{ch}$ (the force from the neighboring beads in the chain), $\vec{f}_n$ (the random force due to the hits of the molecules of solvent), and $\vec{f}_n^{fr}$ (the friction force on the bead during its motion in the solvent). With the hydrodynamic interaction taken into account, the latter force is the Stokes one,

$$\vec{f}_n^{fr} = -\xi \left[ \frac{d\vec{x}_n}{dt} - \vec{v}(\vec{x}_n) \right], \tag{1}$$

where $\vec{x}_n$ is the position vector of the bead and $\vec{v}(\vec{x}_n)$ is the velocity of the solvent in the place of the $n$th bead due to the motion of other beads. The friction coefficient is $\xi = 6\pi\eta b$ with $\eta$ being the solvent viscosity and $b$ the radius of the bead. In Refs. [10, 11], we have generalized the RZ equation by taking into account the viscous aftereffect. Instead of the Stokes force (1), the Boussinesq friction force was used and the Oseen tensor was built on the basis of nonstationary Navier-Stokes equation. Here we give the modified RZ equation after the Fourier transformation in the time. In the continuum approximation with respect to the discrete variable $n$, the equation of motion for the $n$th bead had the form

$$-i\omega x_\alpha^\omega(n) = \frac{1}{\xi^\omega} \left[ f_\alpha^{ch,\omega}(n) + f_\alpha^\omega(n) + M\omega^2 x_\alpha^\omega(n) \right]$$

$$+ \int_0^N dm H_{\alpha\beta nm}^\omega \left[ \frac{3k_B T}{a^2} \frac{\partial^2 x_\beta^\omega}{\partial m^2} + f_\beta^\omega(m) + M\omega^2 x_\beta^\omega(m) \right] \tag{2}$$

In this equation, *a* is the mean square distance between the beads along the chain, *M* is the bead mass,

$$\xi^\omega = \xi\left[1+\chi b+\frac{1}{9}(\chi b)^2\right], \tag{3}$$

is the frequency-dependent friction coefficient with the positive real part of $\chi = \sqrt{-i\omega\rho/\eta}$, $\rho$ is the solvent density, and $H^\omega_{\alpha\beta nm} = H^\omega_{\alpha\beta}(|\vec{x}(n)-\vec{x}(m)|)$ is the Fourier transformation of the Oseen tensor. The explicit expression for this tensor is given in [26]. After the preaveraging of the tensor over the equilibrium pair distribution function

$$P(r_{nm}) = (2\pi a^2|n-m|/3)^{-3/2}\exp\left[-3r_{nm}^2/(2a^2|n-m|)\right], \quad \vec{r}_{nm} \equiv \vec{x}(n)-\vec{x}(m), \tag{4}$$

the result reads

$$\langle H^\omega_{\alpha\beta nm}\rangle_0 = \delta_{\alpha\beta}h^\omega(n-m), \tag{5}$$

$$h^\omega(n-m) = (6\pi^3|n-m|)^{-1/2}(\eta a)^{-1}\left[1-\sqrt{\pi}z\exp(z^2)\mathrm{erfc}(z)\right], \quad z \equiv \chi a(|n-m|/6)^{1/2}.$$

The approximation significantly simplifies Eq. (2), which becomes the system of linear integro-differential equations with the random sources (Langevin equations). In the continuum limit one has to add the conditions for the end beads,

$$\partial\vec{x}^\omega(n)/\partial n = 0, \quad n = 0, N \tag{6}$$

The solution of Eq. (2) can be searched for in the form of Fourier expansion in the internal modes of the chain

$$\vec{x}^\omega(n) = \vec{y}_0^\omega + 2\sum_{p\geq 1}\vec{y}_p^\omega\cos(\pi np/N), \quad \vec{y}_p^\omega = \frac{1}{N}\int_0^N dn\,\vec{x}^\omega(n)\cos(\pi np/N). \tag{7}$$

In this case the amplitudes $\vec{y}_p^\omega$ are determined by linear algebraic equations. The elements of the Fourier transformation of the matrix

$$h^\omega_{pq} = \frac{1}{N^2}\int_0^N dn\int_0^N dm\,h^\omega(n-m)\cos(\pi pn/N)\cos(\pi pm/N) \tag{8}$$

are diagonal at large values of the indexes. If both the indexes are of the order of 1, the non-diagonal elements are much smaller than the diagonal ones [4]. Neglecting the non-diagonal elements, in Eq. (8) the result of the transformation for $p \approx 1$ can be used in the same form as for large *p*. Then we have for the Fourier amplitudes

$$\vec{y}_p^\omega = \vec{f}_p^\omega\left[-i\omega\,\Xi_p^\omega - M\omega^2 + K_p\right]^{-1}, \tag{9}$$

where

$$\Xi_p^\omega = \xi^\omega\left[1+(2-\delta_{p0})Nh^\omega_{pp}\right]^{-1}, \quad K_p = 3\pi^2p^2k_BT/(Na)^2, \; p = 0, 1, 2\ldots \tag{10}$$

The spectral properties of the amplitudes of the random forces $\vec{f}_p^\omega$ follow from the fluctuation-dissipation theorem (FDT) [27] so that

$$\left\langle f_{p\alpha}^{\omega} f_{q\alpha'}^{\omega'} \right\rangle = \frac{k_B T}{(2-\delta_{p0})\,\pi\,N} \operatorname{Re} \Xi_p^{\omega} \delta_{\alpha\beta} \delta_{pq} \delta(\omega+\omega'). \tag{11}$$

Using these properties, we have in the work [11] analyzed the dynamics of the polymers in the dilute solution. This is not the unique way how to determine the spectral properties of these forces in the theory of the Brownian motion. In what follows we shall study the properties of the random forces based on a different approach.

## 2. Effects of hydrodynamic noise

At the motion of a spherical particle in liquid we shall take into account, along with the velocity field created by the motion of this particle, the additional velocity and pressure fields due to the spontaneous fluctuations of the tensor of random stresses, $S_{\alpha\beta}$ (the spontaneous hydrodynamic noise). The noise is assumed to be Gaussian, with the first moment equal to zero, and the quadratic fluctuations of the tensor are defined by the delta-correlated expression [27]

$$\left\langle S_{\alpha\beta}(\vec{r},t) S_{\alpha'\beta'}(\vec{r}',t') \right\rangle = 2 k_B T \eta \left( \delta_{\alpha\alpha'} \delta_{\beta\beta'} + \delta_{\alpha\beta'} \delta_{\alpha'\beta} - \tfrac{2}{3} \delta_{\alpha\beta} \delta_{\alpha'\beta'} \right) \delta(\vec{r}-\vec{r}')\,\delta(t-t'). \tag{12}$$

Let the velocity $\vec{v}^{\omega}(\vec{r})$ and the pressure $p^{\omega}(\vec{r})$ be the Fourier components of the field of hydrodynamic noise, which is created by the random stresses $S_{\alpha\beta}^{\omega}$ in the absence of the particle, and the fields in the presence of the particle moving with the velocity $\dot{\vec{x}}_n$ will be denoted by $\vec{V}^{\omega}(\vec{r})$ and $P^{\omega}(\vec{r})$. The origin of the spherical system of coordinates will be chosen in the center of inertia of the particle. The boundary problem for the determination of the Fourier components of the velocity and pressure is written in the following form:

$$-i\omega\rho\vec{V}^{\omega} = -\nabla P^{\omega} + \eta \Delta \vec{V}^{\omega} + \vec{F}^{\omega}, \quad F_{\alpha}^{\omega} = \nabla_{\beta} S_{\alpha\beta}^{\omega}, \qquad \operatorname{div}\vec{V}^{\omega} = 0 \tag{13}$$

$$\vec{V}^{\omega}(r=b) = \dot{\vec{x}}_n^{\omega}, (|\vec{r}-\vec{x}_n| = b); \qquad \vec{V}^{\omega}(\vec{r}) \to \vec{v}^{\omega}(\vec{r}),\ (r \gg b). \tag{14}$$

The solution of a similar problem is given in the works by Bedeaux and Mazur [28 - 30], where it was used for the determination of the tensor of stresses and the hydrodynamic force acting on the particle. We use these results and represent the hydrodynamic force on the elements of the polymer chain in the form of two contributions. The first of them coincides with the nonstationary expression (1),

$$\vec{f}_n^{fr,\omega} = -\xi^{\omega} \left[ \dot{\vec{x}}_n^{\omega} - \vec{v}^{\omega}(\vec{x}_n) \right], \tag{15}$$

and the second one is expressed as a random force, the properties of which are determined by the correlators (12),

$$\vec{f}_n^{\omega} = \xi \left[ (1+b\chi)\,\vec{v}^{S\omega}(\vec{x}_n) + \tfrac{1}{3} b^2 \chi^2\,\vec{v}^{V\omega}(\vec{x}_n) \right] \tag{16}$$

($\chi$ is defined after Eq. (3)). Here, the following integrals over the surface and volume of the particle are introduced, the center of inertia of the particle being placed in the point $\vec{x}_n$:

$$\vec{v}^{S\omega}(\vec{x}_n) = S^{-1} \int \vec{v}^{\omega}(\vec{x}_n + b\vec{n}_0)\,dS, \quad \vec{v}^{V\omega}(\vec{x}_n) = V^{-1} \int \vec{v}^{\omega}(\vec{x}_n + \vec{r})\,dV. \tag{17}$$

Note that integrating at $\vec{x}_n = 0$ for the bilinear averages of the random force (16) and using (12), the result [30] exactly coincides with the traditional one, based on FDT [27],

$$\langle f_\alpha^\omega f_\beta^{\omega'}\rangle = 2k_B T \operatorname{Re}\xi^\omega \delta_{\alpha\beta}\delta(\omega+\omega'). \tag{18}$$

We use equations (15), (16), and (17) to construct bilinear spectra of the internal amplitudes of the polymer chain,

$$\langle y_{p\alpha}^\omega y_{q\beta}^{\omega'}\rangle = \frac{\langle f_{p\alpha}^\omega f_{q\beta}^{\omega'}\rangle}{(-i\omega\Xi_p^\omega - M\omega^2 + K_p)(-i\omega'\Xi_q^{\omega'} - M\omega'^2 + K_q)} \tag{19}$$

which are connected to the spectral densities of the noise by the integral transformation

$$\langle f_{p\alpha}^\omega f_{q\beta}^{\omega'}\rangle = \frac{1}{N^2}\int_0^N dn \int_0^N dm \langle f_{n\alpha}^\omega f_{m\beta}^{\omega'}\rangle \cos\frac{\pi p n}{N}\cos\frac{\pi q m}{N}. \tag{20}$$

The quadratic fluctuations (12) of the stress tensor are delta-correlated so that one can immediately write

$$\langle f_{n\alpha}^\omega f_{m\beta}^{\omega'}\rangle = \delta(\omega+\omega')\langle \hat{A}\hat{A}' v_\alpha^\omega(\vec{x}_n+\vec{r}) v_\beta^{\omega'}(\vec{x}_m+\vec{r}')\rangle, \quad \vec{f}_n^\omega = \hat{A}\vec{v}^\omega(\vec{x}_n+\vec{r}), \tag{21}$$

where we have introduced the operator $\hat{A}$ acting according to the rule (16). The spectral density of the fluctuations of the velocity field due to the influence of noise is well known and is determined by integrating the hydrodynamic susceptibility [31]

$$\langle v_\alpha^\omega(\vec{R}) v_\beta^{\omega*}(\vec{R}')\rangle = \delta_{\alpha\beta}\frac{k_B T}{12\pi^3\rho}\int \frac{\nu k^2 \exp(i\vec{k}(\vec{R}-\vec{R}'))}{\omega^2+\nu^2 k^4}d\vec{k}, \tag{22}$$

where $\nu$ is the kinematic viscosity. Before performing the double integration in Eq. (20) over the discrete variables (the continuum approximation), we average the exponential in Eq. (22) over the equilibrium distribution function $P(r_{nm})$ (Eq. (4)) of the chain elements (as in the case of the Oseen tensor):

$$\langle \exp(i\vec{k}(\vec{x}_n-\vec{x}_m))\rangle_0 = \int \exp(i\vec{k}\vec{r}_{nm})P(r_{nm})d\vec{r}_{nm} = \exp\left[-\frac{k^2 a^2}{6}|n-m|\right]. \tag{23}$$

From here, after the integral transformation we find in the same approximation as for the Oseen matrix elements $h_{pq}$

$$\left[\langle \exp(i\vec{k}(\vec{x}_n-\vec{x}_m))\rangle_0\right]_{pq} \approx \delta_{pq}\frac{24}{pNa^2}\frac{k^2}{k^4+(6\pi p/Na^2)^2}. \tag{24}$$

Now, Eq. (20) can be written in the form

$$\langle f_{p\alpha}^\omega f_{q\beta}^{\omega'}\rangle \approx \delta(\omega+\omega')\delta_{pq}\delta_{\alpha\beta}\frac{2k_B T}{\pi^3 pNa^2\eta}\int d\vec{k}\,\frac{\hat{A}\hat{A}'^* \exp(i\vec{k}(\vec{r}-\vec{r}'))}{k^4+(\omega/\nu)^2}\frac{k^4}{k^4+(6\pi p/Na^2)^2}. \tag{25}$$

After the action of the operator $\hat{A}$ we have

$$\hat{A}\exp(i\vec{k}\vec{r}) = \xi\left[(1+b\chi)\frac{\sin bk}{bk} + (b\chi)^2\left(\frac{\sin bk}{bk}-\cos bk\right)(bk)^{-2}\right]. \tag{26}$$

Now the spectral density of the amplitudes $y_p^\omega$ can be written for $p = 1, 2, \ldots$ as

$$\left\langle |y_p^\omega|^2 \right\rangle = \frac{24 k_B T b}{\pi^2 p N a^2 \eta} \int_0^\infty dk \, \frac{k^6}{k^4 + (\omega b^2 / \nu)^2} \, \frac{1}{k^4 + (6\pi p b^2 / N a^2)^2}$$

$$\times \left| (1+b\chi)\frac{\sin k}{k} + (b\chi)^2 \left( \frac{\sin k}{k^3} - \frac{\cos k}{k^2} \right) \right|^2 \left| \frac{\xi}{-i\omega \Xi_p^\omega - M\omega^2 + K_p} \right|^2. \tag{27}$$

The integration over the dimensionless variable can be performed using the decomposition of the integrand in simple fractions. The answer can be expressed in terms of elementary functions and the error function. It has a bulky form and we shall not give it here. Instead, we shall consider some more simple consequences of the theory.

To investigate the diffusion motion of the polymer coil as a whole one should analyze the dynamical properties of the radius vector of the center of inertia of the polymer, $\vec{y}_0^\omega = \frac{1}{N}\int_0^N dn \, \vec{x}^\omega(n)$. The diffusion coefficient of the coil can be found using the Kubo formula

$$D = \frac{1}{3}\int_0^\infty \langle \dot{\vec{y}}_0(t)\dot{\vec{y}}_0(0)\rangle dt, \tag{28}$$

Equivalently, it can be expressed through the spectral density $\left\langle |\dot{y}_0^\omega|^2 \right\rangle \Big|_{\omega=0}$. Using Eqs. (9), (10), and (16), we obtain

$$\left\langle |\dot{y}_0^\omega|^2 \right\rangle = \frac{k_B T}{\pi^2 N^2 \rho} \int_0^N dn \int_0^N dm \int_0^\infty dk \, \frac{\nu k^4}{\omega^2 + (\nu k^2)^2} \left( \frac{\sin kb}{kb} \right)^2 \exp(-k^2 a^2 |n-m|/6). \tag{29}$$

The diffusion coefficient can be easily obtained from the Kubo relation when $b = 0$:

$$D_Z = \frac{k_B T}{3\pi^2 N^2 \eta} \int_0^N dn \int_0^N dm \int_0^\infty dk \, \exp(-k^2 a^2 |n-m|/6) = \frac{k_B T}{\sqrt{6\pi^3} N^2 \eta a} \int_0^N dn \int_0^N \frac{dm}{\sqrt{|n-m|}}. \tag{30}$$

The double integral equals to $8N^{3/2}/3$ and we get the correct diffusion coefficient in the Zimm model [3],

$$D_Z = \frac{8 k_B T}{3(6\pi^3 N)^{1/2} a\eta}. \tag{31}$$

At $b > 0$ we introduce the dimensionless parameter $\sigma^2 = 6b^2/(a^2 N) = b^2/R_G^2$ ($R_G$ is the gyration radius of the polymer) and use

$$\int_0^N dn \int_0^N dm \, f(|n-m|) = \int_0^N dn \left\{ \int_0^n ds \, f(s) + \int_0^{N-n} ds \, f(s) \right\} = 2\int_0^N (N-s) f(s) ds.$$

For $f(s) = \exp(-k^2 s/\sigma^2 N)$, the double integral from 0 to $N$ in Eq. (29) equals

$$\frac{2\sigma^2 N^2}{k^2}\left(1 - \frac{\sigma^2}{k^2} + \frac{\sigma^2}{k^2}\exp(-k^2/\sigma^2)\right).$$

Then the diffusion coefficient is expressed as

$$D = D_Z \Psi(\sigma), \qquad \Psi(\sigma) = \frac{3}{2\sqrt{\pi}} \int_0^\infty \frac{dx}{x^4} \left(\frac{\sin \sigma x}{\sigma x}\right)^2 \left(x^2 - 1 + e^{-x^2}\right), \tag{32}$$

with the limiting value $\Psi(0) = 1$, when $D = D_Z$. Integrating Eq. (32), the function $\Psi$ can be expressed in terms of elementary functions and the error function. The effective hydrodynamic radius of the coil determined from the Einstein relation $R_C = k_B T/(6\pi \eta D)$ is $R_C \sim a\sqrt{N}/\psi(\sigma)$ (see Eq. 31). It thus contains a weak dependence on the ratio between the size of the bead $b$ and the size of the coil $a\sqrt{N}$.

Let us also consider the VAF of the coil,

$$\Phi_0(t) = \langle \dot{\vec{y}}_0(t) \dot{\vec{y}}_0(0) \rangle = \frac{1}{2\pi} \int d\omega \cos \omega t \left\langle \left|\dot{\vec{y}}_0^\omega\right|^2 \right\rangle. \tag{33}$$

At long times it is sufficient to restrict ourselves to small $\omega$. Again, the most simple case corresponds to $b = 0$. Equations (19) - (27), using the found double integral over $n$ and $m$ and the integration over $\omega$, yield

$$\Phi_0(t) = \frac{6k_B T}{\pi^2 \rho N a^2} \int_0^\infty dk \exp(-k^2 \nu t) \left(1 - \frac{6}{Na^2 k^2} + \frac{6}{Na^2 k^2} \exp(-k^2 N a^2/6)\right). \tag{34}$$

The result of integration is

$$\Phi_0(t) = \frac{3k_B T}{\pi^{3/2} \rho Na^2} \left\{ \frac{1}{\sqrt{\nu t}} + \frac{12}{Na^2} \left[\sqrt{\nu t} - \sqrt{\nu t + Na^2/6}\right] \right\}. \tag{35}$$

It gives the correct asymptote at long times $t \gg Na^2/6\nu$, which is independent on any polymer parameters [22, 23],

$$\Phi_0(t) \approx \frac{k_B T}{4\rho(\pi \nu t)^{3/2}}. \tag{36}$$

This expression is the same as for individual Brownian particles when the viscous aftereffect is taken into account. At short times $\nu t \ll Na^2/6$ (assuming however $t \gg b^2 \rho/\eta$) we find for the Zimm polymer $\Phi_0(t) \sim t^{-1/2}$

$$\Phi_0(t) \approx \frac{3k_B T}{Na^2} \frac{1}{\sqrt{\pi^3 \rho \eta t}}. \tag{37}$$

For arbitrary $b$ the VAF (33) is instead of Eq. (34) expressed through the integral

$$\Phi_0(t) \propto \int_0^\infty dk \left(\frac{\sin kb}{kb}\right)^2 \exp(-k^2 \nu t) \left(1 - \frac{1}{R_G^2 k^2} + \frac{1}{R_G^2 k^2} \exp(-k^2 R_G^2)\right). \tag{38}$$

The integral can be evaluated in terms of the error function and elementary functions. At $b = 0$ we return to the previous result (35).

Knowing the VAF $\Phi_0(t)$, it is straightforward to calculate the MSD of the polymer coil. It is linked with $\Phi_0(t)$ by means of the relation [16]

$$\langle \Delta \vec{r}^2(t) \rangle = 2 \int_0^t (t - t') \Phi_0(t') dt'. \tag{39}$$

The calculation using Eq. (35) is easy but leads to a rather bulky expression. The result for long times $t \gg Na^2/6\nu$ that corresponds to Eq. (36) has the form

$$\langle \Delta \vec{r}^2(t) \rangle \approx \frac{8k_BT}{(6\pi^3 N)^{1/2} a\eta} \left[ t - \left(\frac{3N\rho}{32\eta}\right)^{1/2} t^{1/2} + ... \right] = 3D_Z \left[ t - \frac{2}{\sqrt{\pi}} (\tau_R t)^{1/2} + ... \right], \quad (40)$$

where the characteristic time $\tau_R = R^2\rho/\eta$ can be expressed through the hydrodynamic radius of the Zimm coil. The second equality in Eq. (40) is written exactly in the form familiar in the theory of the Brownian motion of rigid particles. Note that the MSD found in Ref. [23] must be corrected: at long times the main terms of the MSD are the Einstein term $\sim t$ and the longest-lived tail, which is negative and proportional to $t^{1/2}$ instead of the positive $\sim t^{3/2}$ [23]. Figure 1 in Ref. [23] also does not correspond to the correct solution: with growing $t$ the MSD should slowly approach the Einstein limit valid for infinitely long times.

## 3. Conclusion

At scales of nanometers to micrometers, typical for long polymers in solution, thermal fluctuations cannot be neglected, but must be incorporated as random terms in the equations of hydrodynamics. Such terms are responsible for the mechanical and thermal energy equations underlying Brownian motion. Since the Brownian motion plays the crucial role in the theory of the dynamics of polymers in solution, the basic equations describing the polymer motion should be the fluctuating hydrodynamics equations. In the present work such approach was used to describe the diffusion of individual flexible polymer coils within the Zimm model with the hydrodynamic interactions between the polymer segments. In spite of rather involved calculations, the final results are quite simple. The obtained expressions significantly differ from the known ones that are based on the Einstein theory of the Brownian motion and the usual stationary Navier-Stokes hydrodynamics for large volumes. The simplest result found for the long-time tail of the velocity autocorrelation function of the coil corresponds to the expression known from the hydrodynamic theory of the Brownian motion of rigid particles and was derived earlier also for the polymers [22, 23]. For the times $t \gg R^2\rho/\eta$ the VAF scales as $t^{-3/2}$. In our theory this law follows from the simple equation (35), which allows obtaining the time dependence of the VAF also for short times, when it scales as $t^{-1/2}$. We have also calculated the mean square displacement and found the diffusion coefficient of the coil in the form of the Zimm diffusion coefficient multiplied by a function that depends on the ratio between the radius of the bead and the size of the whole coil. The theory for the coil diffusion can be supplemented by the consideration of the internal motion of the polymer. It could be done but, as shown in Refs. [11, 20], the effects of the nonstationary hydrodynamics on the relaxation of internal modes are weak and hardly detectable experimentally. As opposite, the "anomalous" diffusion of the polymers at short times should be observable in dynamic scattering experiments.

**Acknowledgment**. This work was supported by the Scientific Grant Agency of the Slovak Republic (VEGA), grant No. 1/3033/06.

**References**

 [1] Rouse P E, 1953 *J. Chem. Phys.* **21** 1272
 [2] Zimm B H, 1956 *J. Chem. Phys.* **24** 269
 [3] Doi M and Edwards S F, 1986 *The Theory of Polymer Dynamics* (Oxford: Clarendon)
 [4] Grosberg A Yu and Khokhlov A R, 1989 *Statistical Physics of Macromolecules* (Moscow: Nauka)
 [5] Richter D, Binder K, Ewen B, and Stühn B, 1984 *J. Phys. Chem.* **88** 6618
 [6] Stockmayer W H and Hammouda B, 1984 *Pure Appl. Chem.* **56** 1372
 [7] Harnau L, Winkler R, and Reineker P, 1996 *J. Chem. Phys.* **104** 6355
 [8] Sawatari N, Yoshizaki T, and Yamakawa H, 1998 *Macromolecules* **31** 4218
 [9] Osa M, Ueda H, Yoshizaki T, and Yamakawa H, 2006 *Polymer J.* **38** 153
[10] Tothova J, Lisy V., and Zatovsky A V, 2003 *J. Chem. Phys.* **119** 13135
[11] Lisy V, Tothova J, and Zatovsky A V, 2004 *J. Chem. Phys.* **121** 10699
[12] Boussinesq J, 1885 *C. R. Acad. Sci. Paris* **100** 935
[13] Vladimirsky V V and Terletzky Ya P, 1945 *Zhur. Eksp. Teor. Fiz.* **15** 258
[14] Giterman M Sh and Gertsenshtein M E, 1966 *Zhur. Eksp. Teor. Fiz.* **50** 1084
[15] Zatovsky A V, 1969 *Izvestia VUZ Fizika* **10** 13
[16] Schram P P J M and Yakimenko I P, 1998 *Physica A* **260** 73
[17] Weitz D A, Pine D J, Pusey P N, and Tough R J A, 1989 *Phys. Rev. Lett.* **16** 1747
[18] Lukić B, Jeney S, Tischer C, Kulik A J, Forró L, and Florin E L, 2005 *Phys. Rev. Lett.* **95** 160601
[19] Lukić B, Jeney S, Sviben Ž, Kulik A J, Florin E L, and Forró L, 2007 *Phys. Rev. E* **76** 011112
[20] Tothova J, Brutovsky B, and Lisy V, 2004 *Laser Phys.* **14** 1511
[21] Lowe C P, Bakker A F, and Dreischor M W, 2004 *Europhys. Lett.* **67** 397
[22] Jones R B, 1980 *Physica A* **101** 389
[23] Bonet Avalos J, Rubí J M, and Bedeaux D, 1991 *Macromolecules* **24** 5997
[24] Brinkman H C, 1947 *Physica* **13** 447
[25] Debye P and Bueche A M, 1948 *J. Chem. Phys.* **16** 573
[26] Zatovsky A V and Lisy V, 2001 *arXiv:cond-mat*/0111465; 2003 *J. Mol. Liq.* **105** 289
[27] Lifshitz E M and Pitaevskii L P, 1978 *Statistical Physics, Part II* (Moscow: Nauka)
[28] Mazur P and Bedeaux D, 1974 *Physica* **76** 235
[29] Mazur P and Bedeaux D, 1974 *Physica* **78** 505
[30] Bedeaux D and Mazur P, 1974 *Physica* **76** 247
[31] Fisher I Z, 1971 *Zhur. Eksp. Teor. Fiz.* **61** 1647